\documentclass[%
reprint
]{revtex4-1}

\usepackage{graphicx}
\usepackage{dcolumn}
\usepackage{bm}
\usepackage{color}
\usepackage{amsmath}
\usepackage{comment}
\usepackage[utf8]{inputenc}
\usepackage{textcomp}
\usepackage{soul}
\usepackage{hyperref}
\usepackage{xr}

\usepackage[symbol]{footmisc}
\renewcommand{\thefootnote}{\fnsymbol{footnote}}

\usepackage{subfiles}
\usepackage{blindtext}
\usepackage{amsmath,amsfonts,amsthm,bm} 
\usepackage{todonotes}

\usepackage{float}
\newfloat{suppfig}{tbh}{losf}
\floatname{suppfig}{Supplementary Figure}

\begin{document}


\title{3D imaging from multipath temporal echoes}%

\author{Alex Turpin$^{1,\dagger,*}$, Valentin Kapitany$^{2,\dagger}$, Jack Radford$^2$, Davide Rovelli$^2$, Kevin Mitchell$^2$, Ashley Lyons$^2$, Ilya Starshynov$^2$, Daniele Faccio$^{2}$} 

\email{Corresponding author: alex.turpin@glasgow.ac.uk / daniele.faccio@glasgow.uk}

\affiliation{$^1$ School of Computing Science, University of Glasgow, Glasgow G12 8QQ, UK\\
$^2$ School of Physics \& Astronomy, University of Glasgow, Glasgow G12 8QQ, UK
}%

\date{\today}
             
\begin{abstract}
Echo-location is a broad approach to imaging and sensing that includes both man-made RADAR, LIDAR, SONAR and also animal navigation. However, full 3D information based on echo-location requires some form of scanning of the scene in order to provide the  spatial location of the echo origin-points. Without this spatial information, imaging objects in 3D is a very challenging task as the inverse retrieval problem is strongly ill-posed. Here, we show that the temporal information encoded in the return echoes that are reflected multiple times within a scene is sufficient to faithfully render an image in 3D. Numerical modelling and an information theoretic perspective prove the concept and provide insight into the role of the multipath information. We experimentally demonstrate the concept by using both radio-frequency and acoustic waves for imaging individuals moving in a closed environment.
\end{abstract}

\maketitle

\begin{figure}[t]
    \centering
    \includegraphics[width=1\linewidth]{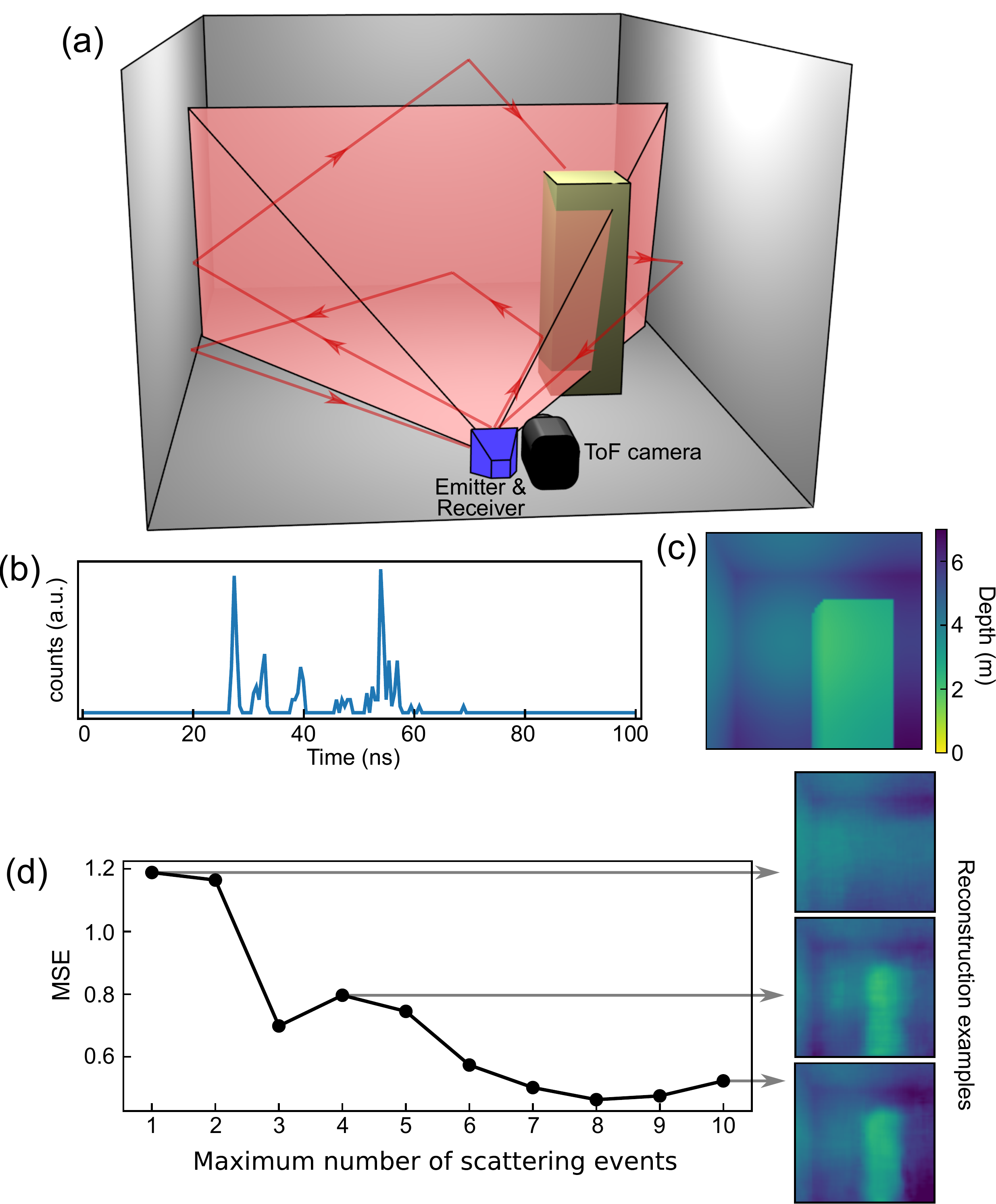}
    \caption{(a) 3D visualisation of our physical system: a rectangular cuboid (yellow) moves within a room. Rays are emitted within a pyramid-like volume and illuminate the scene. Red arrows indicate examples of multipath reflections, which eventually reach the detector (blue) that records their arrival time. (b) An example of a recorded time histogram. (c) Color-depth encoded 3D view of the scene. (d) Mean mean-squared-error (MSE) with increasing multipath contributions, calculated between the ground truth 3D scene and the neural network reconstruction, averaged over 100 3D images. Insets show depth image reconstruction examples obtained for 1-path, 4-path, and 10-path events.}
    \label{fig1}
\end{figure}

\textit{\textbf{Introduction.}}
In nature, detecting and locating objects from reflected echoes is generally possible only if two or more detectors are used. 
Animals such as bats or dolphins \cite{bats_dolphins} and even humans \cite{human_clicks} can emit pulses of sound to sense the environment they navigate through and identify objects. RADAR and LiDAR imaging systems operate in a similar way, albeit with electromagnetic (EM) radiation (radio waves and light, respectively): a series of EM pulses are used to scan and probe the scene and, by measuring the arrival time of the return echoes and correlating this with the direction from which they are detected, they can form a 3D estimate of the scene \cite{book-lidar,skolnik:1980}. This principle also holds for non-line-of-sight (NLOS) applications \cite{2020:natrevphys:faccio,wetzstein:2018:nature,2019:nature:velten,Metzler:2020:optica,faccio:2017:oe}, where photon echoes of light, now scattered from multiple surfaces along indirect paths, are analysed with the goal of revealing the 3D shape and visual appearance of objects outside the direct line of sight. Although NLOS is typically deployed with optical sources, it has also been demonstrated with acoustic \cite{Lindell:2019:Acoustic} and radio-frequency (RF) sources \cite{heide:2020:CVPR}. \\
Locating objects in space and forming an image in 3D from their wave echoes using a single point detector without any form of scanning is, computationally-speaking, a strongly ill-posed problem and therefore considerably more challenging. However, recent work has shown that echoes contain a very rich structure in the time dimension that can be used to extract meaningful information about the scene \cite{caramazza:2018:scirep,metzler:2019:keyhole,turpin:2020:optica}. 
In these cases, further assumptions of the scene are required in order eliminate ambiguities arising from the fact that the echo is single-path, i.e. the outgoing signal reflects only once from the scene objects. This leads to ambiguity in the form of an equal-distribution-probability for the echo origin point that is spread over a spherical dome centred on the detector and with a radius determined by the echo arrival time. 
The additional assumptions referred to above can be introduced, for example, in the form of additional information by means of a machine learning algorithm that exploits the knowledge of static objects in the scene background and a statistical knowledge of the objects that we want to image  \cite{caramazza:2018:scirep,turpin:2020:optica}. \\
The paradigm investigated here is the extension of echo detection to multipath trajectories of the return signal. 
The idea of using multipath reflections for sensing inside buildings, through walls or out of view, especially with RF waves, has been a topic of extensive study during the last decade \cite{krolik:2006,setlur:2011,sen:2010,leigsnering:2014,muqaibel:2015,fuschini:2017,goulianos:2017,neubauer:1986}. 
However, these simple geometric approaches are typically limited to locating the position of objects (and not imaging), e.g. of humans inside known environments.  Multipath sensing has also been combined with Bayesian inference \cite{lim:1996} and convolutional neural networks \cite{ferguson:2018} to localise sonic sources. 
In the optical domain, multipath interference, i.e. the contribution from light following multiple paths onto the same pixel, is generally considered problematic and has to be accounted for to acquire accurate depth maps \cite{bhandari:2014,freedman:2014,goyal:2016,marco:2017}. However, recent works have explored multipath optical sensing both theoretically \cite{gkioulekas2017} and experimentally \cite{nam2020} by exploiting deterministic algorithms that provide mathematical proof for the ability to reconstruct the geometry of simple scenes from a single location. \\
In this work, we provide empirical evidence that 3D scenes can be reconstructed from temporal echoes alone.   We make use of a data-driven approach that exploits multipath temporal echoes, i.e. echoes from waves that are reflected multiple times from surfaces and objects within a scene, to unambiguously reconstruct a meaningful 3D image in a fixed scenario. We first present numerical simulations that show how a simple artificial neural network can be trained to reconstruct a 3D scene. We then underline the importance of the multipath echoes, with a dominant role played by the first few reflections and a gradually decreasing importance of further bounces. These findings are supported by an information theoretic analysis applied to the raw multipath data that is independent of the image retrieval algorithm. We then demonstrate our approach experimentally. Although our method could be in principle implemented with optical pulses, light suffers from severe diffused reflection, which would make it very hard to detect any optical signal after 2 reflections. We therefore concentrate on GHz EM radio-frequency (RF) and kHz acoustic waves, as these can be reflected multiple times by walls and objects. In both cases, we are able to precisely retrieve 3D images of a dynamic scene with a significant improvement beyond what is achievable using single-path echoes.\\ 
 \textit{\textbf{3D imaging with multipath temporal echoes.}}
Our approach is conceptually sketched in Fig.~\ref{fig1}. A source emits waves in the form of pulses that diverge with a wide angle so as to flash-illuminate the whole scene. 
The emitted pulses are then reflected by the room walls and the objects inside it and, finally, are detected by a single-pixel sensor with time-resolving capabilities. The timing of successive pulses is arranged so as to not temporally overlap with any returning echoes, i.e. each outgoing pulse and detection of return echoes are completely separate events from the emission of a successive pulse.
The sensor collects and records the received energy over a wide angle and provides this information in the form of a temporal histogram. The process of pulsed waves bouncing multiple times inside the room is fully deterministic: with a complete knowledge of the distribution of objects within the room, the room dimensions, and their reflectivity, it is straightforward to predict the recorded temporal histogram. However, solving the inverse process, namely the reconstruction of the scene (including room and objects) in 3D dimensions from just the temporal histogram, is ill-posed: echoes arriving to the detector at time $t_d$ are compatible with objects placed not just at a single point (as would be desired), but rather with the whole surface of a spherical dome represented by the equation $(ct_d)^2/2 = x^2+y^2+z^2$ (where $c$ is the speed of the pulse). This ambiguity has been previously solved, although only in part, by utilising the fact that a moving object will obscure static background objects, therefore removing them from return echo patterns \cite{turpin:2020:optica}.\\
In contrast, in this work we highlight the strength of including multipath reflections in the data-driven solution to solve the ambiguity issue: using not only the first reflection but 2, 3, and more reflections breaks the degeneracy and helps the algorithm to reconstruct the position and shape of the object in 3D with high accuracy, thus making background objects not essential.\\ 
 \textit{\textbf{Numerical simulations.}}
We first show numerical simulations based on Monte Carlo ray-tracing (see \cite{SM} for full details). Our scene consists of a closed room with walls, floor and ceiling that all have the same 100\% reflectivity [Fig.~\ref{fig1}(a)]. Inside this room, a rectangular cuboid is placed in different positions and the scene is imaged in 3D with a ToF camera providing a 2D depth map, see Fig.~\ref{fig1}(c). We consider that an emitter emits probe pulses in all directions within azimuth and elevation angles $\theta$ and $\phi$, both within $[-67.5^{\circ},67.5^{\circ}]$. The return echo amplitudes, i.e. the number of returning rays per time [Fig.~\ref{fig1}(b)], are recorded in time at the detector that is co-located with the emitter. 
Each scene is sampled with 10000 rays per object position, for 2000 objects positions. 
This provides a data set of temporal trace-3D image pairs that we use to train a convolutional deep neural network, shaped such as to force information through a bottleneck (see \cite{SM} for details) to extract features from data. We then test the neural network with histograms that were never used during training and render an estimate of the scene in 3D. We repeat this analysis for an increasing number of path events, starting from single-path until 10-path events, and we analyse the quality of the reconstructions in terms of the mean-square error (MSE) between the ground truth and the retrieved images (see \cite{SM,kingma2014adam,tensorflow2015,keras} for further details). To avoid specificity of the training by the deep neural network architecture, we re-train the network 10 times for each path event, such that for every training round we leave the starting weights of the neural network random. This procedure guarantees a slightly different image reconstruction every time the algorithm is trained. Then, we average our reconstruction-quality metrics over these 10 networks. Our results, summarised in Fig.~\ref{fig1}(d), show that the MSE decreases as the number of multipath events is increased. In particular, we see that the first 2-4 multipath echoes are the most important and significantly improve scene reconstruction. This can be seen clearly not only in the MSE but also in the insets to Fig.~\ref{fig1}(d) that show examples of a reconstruction for 1, 4 and 10 path events. We clearly see that whilst for single-path it is hard to distinguish the object position due to blurring arising from the above mentioned ambiguities, multipath information cures this problem and allows to clearly resolve the 3D scene (see \cite{SM} for further examples). 
We quantify the gain in information when including an increasing number of paths using the concepts of Shannon entropy, mutual information and joint entropy as derived in Information Theory \cite{shannon, MacKay_info, elements_of_info}. The Shannon entropy  gives the expectation value of uncertainty reduction when observing a variable $X$ at values $x_i$, which occur with probability $p(x_i)$:
\begin{equation}\label{Eq:entropy}
    H(X) = -\sum_{i=1}^N{p(x_i)\log_2{p(x_i)}}.
\end{equation}
More specifically, we take a set of 2000 examples of individual temporal histograms from the numerical model described above, within which we identify histogram shapes $x_i$ that occur with probability $p(x_i)$. We can then calculate the joint entropy $H({X},{Y})$ for single-path histograms ${X}$ and 2-path histograms ${Y}$: 
\begin{equation}\label{Eq:joint_entropy}
    H({X},{Y}) = -\sum_{j=1}^M{\sum_{i=1}^N{p({x}_i, {y}_j)\log_2{p({x}_i,{y}_j)}}}.
\end{equation}
This can  be extended to calculate the joint entropy for data containing $<n$ bounces and $<(n+1)$ bounces.
The mutual information, $MI({X};{Y})$, then describes the information shared by the two random variables due to correlations within the data:
\begin{equation}\label{Eq:mutual_info}
    MI({X};{Y}) = H({X})+H({Y})-H({X},{Y}).
\end{equation}
We rearrange Eq.~(\ref{Eq:mutual_info}) to find the additional \emph{uncorrelated} information, $UI$, in the multipath data ${Y}$, i.e. the mutual information $MI({X};{Y})$ subtracted from the total information, $H({Y})$. In other words, the additional information that is gained by including photons from a second or multiple reflections/paths is given by $UI({X};{Y})=H({X},{Y})- H({X})$.\\
\begin{figure}[h!]
    \centering
    \includegraphics[width=1\linewidth]{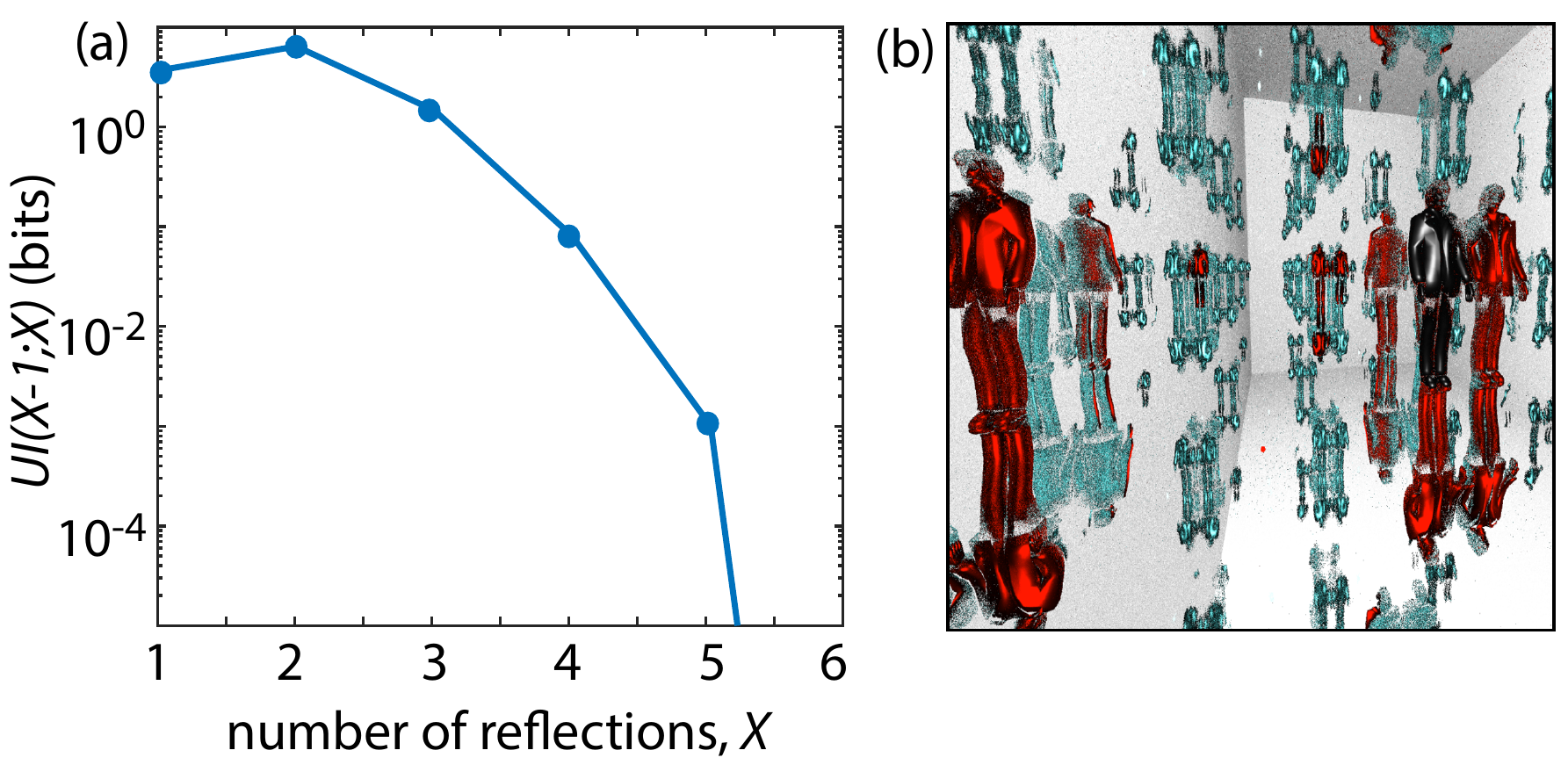}
    \caption{(a) The gain in information when including photons in the temporal data which have experienced an increasing number of reflections within the scene. (b) A simulation of a multipath scene as would be viewed by a camera. The various reflections show different viewpoints of the mannequin therefore intuitively explaining why multipath echoes contain additional information but also why beyond the $4^{th}$ bounce, there is little or no gain of information (see text for  details).}
    \label{fig:joint_entropy}
\end{figure}
Figure~\ref{fig:joint_entropy}(a) shows $UI({X-1};X)$ in log scale for increasing number of reflections/paths. As can be seen, significant additional (uncorrelated) information is gained from the 2nd and 3rd reflections but becomes negligible after 4 reflections. Remarkably, in this configuration $UI$ for a 2-path signal is larger than the information contained in the direct 1-path (standard LIDAR, single reflection) signal. An intuitive insight into understanding this gain in information from multipath data is shown in Fig.~\ref{fig:joint_entropy}(b): the 3D dimensional rendering of a scene, as would be observed by a camera placed at the detection point, appears very similar to what would be observed in a room of mirrors. The first reflection (in black) provides only direct line-of-sight information of the object; the first 4 reflections (in red) show different effective viewpoints (side-view and back-view) that would otherwise be inaccessible and therefore increase the information; all successive reflections (in light blue) are replicas of the first 4 reflections and do not contain additional useful information. That said, we underline that in real life scenarios, the noisy-channel coding theorem \cite{shannon} indicates that adding redundant replicas of information in the form of higher order paths, could still lead to preservation of information that is lost due e.g. to measurement noise.\\
\begin{figure*}[t!]
    \centering
    \includegraphics[width=0.87\linewidth]{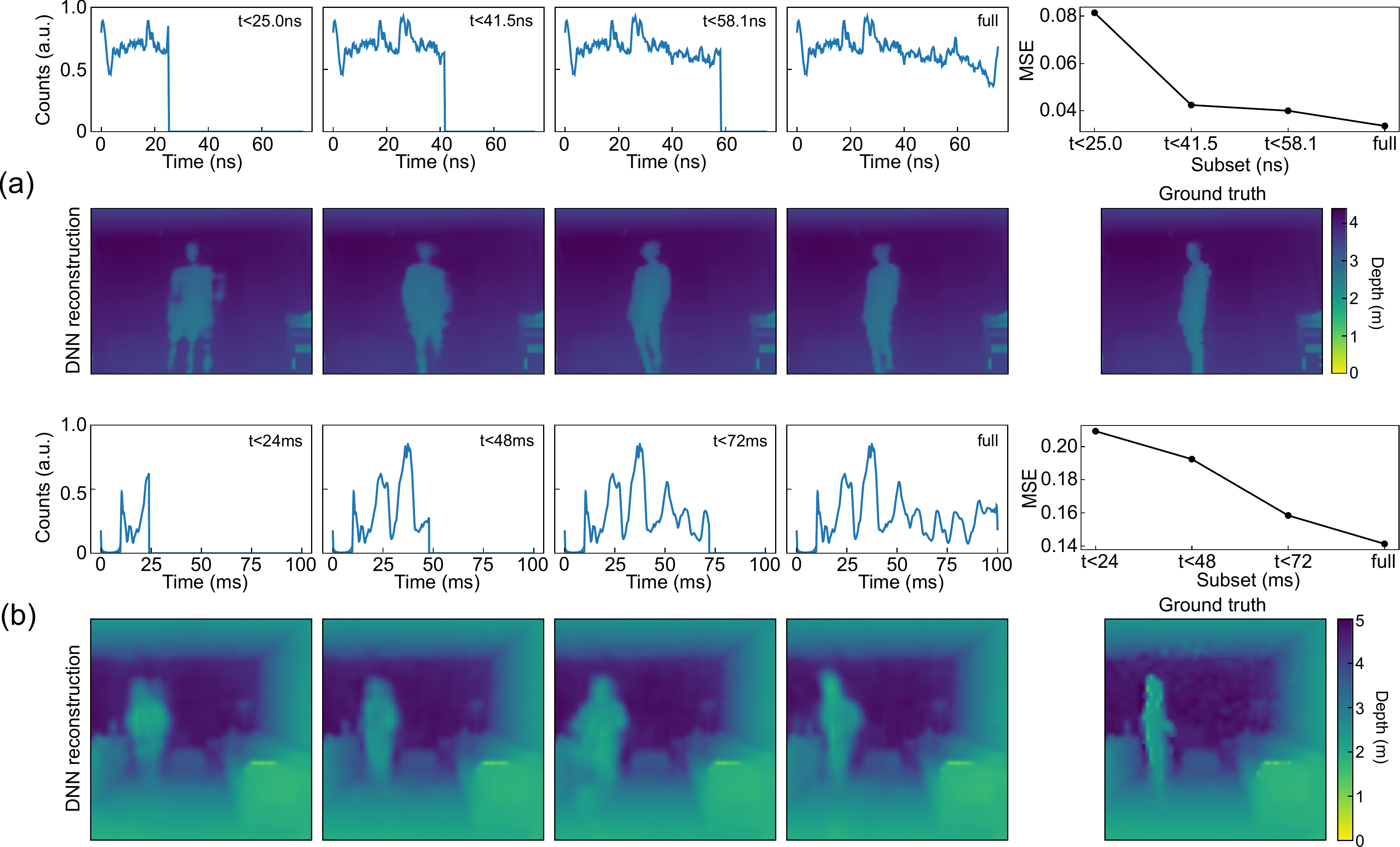}
    \caption{(a) RF and (b) acoustic results. The top rows of (a) and (b) show the time histograms that are truncated at increasing times, therefore including an increasing number of multipath echoes. The last plot of first rows show the quality of the image reconstructions in terms of mean-squared error (MSE) compared to the ground truth for a set of 100 scenes, for increasing multipath events. The second row in (a) and (b) shows the corresponding images retrieved with the deep neural network, and the ToF camera ground truth image.}
    \label{fig:experiments}
\end{figure*}
 \textit{\textbf{Experiments.}}
We show the validity of our approach with experiments using two different sources of  waves, namely GHz radio-frequency (RF) and kHz-frequency acoustic pulses. The experimental set-up in both cases is identical to Fig.~\ref{fig1}(a), where the emitter/detector is an RF-antenna or a speaker and microphone, for the RF and acoustic experiments, respectively. \\
For the experiments with RF waves, we use a transceiver module (TI-AWR1642), which operates in the frequency modulated continuous wave regime \cite{TI_radar}, with a range resolution and maximal unambiguous range of 4.4~cm and 9~m respectively. The transceiver probes the scene with an angular aperture of 20$^\circ$ in the vertical plane and 180$^\circ$ in the horizontal plane (-3~dB FWHM). An analog-to-digital converter samples the signal with 120~ns temporal resolution and 133 Hz rate.\\
The experiments were conducted with a human individual walking around in a room  with approximate dimensions of $3\times4\times2.5 \,\rm{m^3}$. The echo recordings from the RF antenna are acquired in parallel to 3D (ground truth) images via a ToF camera (Basler), which provides $80\times60$ pixel color-encoded depth images. \\
For the acoustic measurements, we replace the RF antenna with a PC speaker (Logitech Z333 system, consisting of 2 speakers + 1 subwoofer) and a PC microphone (integrated in a Logitech C270 webcam). The speakers emit a pulsed wave with centre frequency of $5\,\rm{kHz}$ ($\lambda \approx 6.7\,\rm{cm}$) and a bandwidth of $1\,\rm{kHz}$, with duration of $50\,\rm{ms}$ and repetition rate $\approx 10\,\rm{Hz}$. The microphone, co-located with the speakers, records the returning echoes for $100\,\rm{ms}$ at a sampling rate of $192\,\rm{kHz}$. The data is Fourier filtered so as to select only signals at $(5 \pm 0.5)\,\rm{kHz}$. The ToF 3D camera used to train the deep learning algorithm was an Intel Realsense D435 capturing $64\times64$ color-encoded depth images. The room used for this experiment had dimensions $7 \times 6 \times 2.5\,\rm{m^3}$). Note that the recording time window in both cases, respectively of $80\,\rm{ns}$ and $100\,\rm{ms}$ for the RF and acoustic experiments, is long enough to ensure that the waves can reach the furthest corner of the rooms and return to the detector.\\
For both the RF and acoustic measurements, we use the pairs of ground truth ToF images and RF (or acoustic) echo temporal traces to train a deep learning algorithm based on convolutional layers followed by a Rectified Linear Unit activation function (see \cite{SM} for details). We use 9000 and 5000 pairs of data for training the neural networks for RF and acoustic data respectively, after which, full 3D images can be retrieved from a single (previously unseen) RF (or acoustic) temporal trace. \\
Figures~\ref{fig:experiments}(a) and (b) show the results for the RF and acoustic cases, respectively (see also \cite{SM,SV} for videos). To explore the role of multipath events, we trained and tested our neural network with successively increased temporal extension of the time histograms: truncation of the data at short times corresponds to single path data, calculated as the ToF to the farthest wall in the room. We increase the truncation time (indicated in the figures) by evaluating the longest ToF value for 2-path and 3-path events in the room so as to include 2 and 3 bounces, thus gradually increasing the information from higher order path contributions. The retrieved 3D scenes [second row in Figs.~\ref{fig:experiments}(a) and (b)] show that networks trained solely on 1-path events [first column of Figs.~\ref{fig:experiments}(a) and (b)] struggle to provide a sharp 3D image as there are many possible scenes that correspond to the same single-path time histogram. Increasing the number of multipath events provides an increasingly improved reconstruction. This improvement can be quantified  by calculating the MSE between the retrieved image and the ground truth, averaged over 865 and 500 different measurements, for RF and acoustics respectively. The MSE in Figs.~\ref{fig:experiments}(a) and (b) (far-right graph), decreases monotonically with increasing multipath contributions, in good agreement with our modelling and experimentally shows the significant 3D imaging capability achieved with multipath temporal echoes. Note that our technique can exploit training on a single individual to operate successfully on different individuals, recovering general shape and position, see \cite{SM}. In this work we focused only on imaging human individuals. Evidence from other work suggests that training with additional objects and geometrical shapes should also be possible \cite{turpin:2020:optica} and generic imaging functionality has been shown in a different but related multipath setting \cite{heide:2020:siggraph}.\\
\indent \textit{\textbf{Conclusions.}}
In summary, we have shown that multipath temporal echoes and deep learning can be used to provide full 3D images of a scene. 
Applications of these ideas might be found in imaging in closed environments so as to enable efficient generation of multipath echoes, for example with healthcare applications for homes and hospitals of the future. Interesting developments might include the generalisation to dynamic background scenarios, to open-air scenes, and to scenes incorporating information from different viewpoints, thus opening applications in NLOS imaging and 3D mapping of complex object geometries. More in general, multipath echo imaging offers interesting opportunities, considering that RF antennas can also be extremely compact (and are currently present in cell phones) and that the acoustic results were obtained with standard computer speakers and microphones, thus effectively transforming everyday household items into full 3D imaging systems. 

\noindent \textbf{Acknowledgements}. We thank Hanoz Bhamgara, Mark Jarvis, and Marton Szafian for technical support with the RF system. D.F. acknowledges financial support from the Royal Academy
of Engineering  and from EPSRC (UK, grant no. EP/T00097X/1). V.K. acknowledges funding from Horiba. A.T. acknowledges support as an LKAS Fellow. Data relevant to this work are available for download at Ref. \cite{UofG_DOI}.
\\ 
$\dagger$ A.T. and V.K. contributed equally to this work.

\newpage

\cleardoublepage

\section{Supplementary Information}

This section contains supplementary material information for the manuscript \textit{3D imaging from multipath temporal echoes}. The document is organised as follows: in Section~\ref{echolocation} we demonstrate through a simple mathematical model that it is possible to locate the position of an object from the multipath temporal echo. In Section~\ref{physical_model} we describe the physical model that we use to validate our approach, while full details on the numerical simulations can be found in Section~\ref{simulations}. Section~\ref{performance_metrics} is devoted to describe the metrics used to analyse our approach. In Section~\ref{algorithm} we describe the deep neural network used for image retrieval and give details of the training. Finally, Section~\ref{informaton} provides full details on the calculations we used to estimate the information carried by multipath echoes. The associated supplementary videos can be found in this \hyperlink{youtube_playlist}{link}.

\subsection{Localisation with multipath temporal echoes}
\label{echolocation}
\begin{figure}[h]
    \centering
    \includegraphics[width=1\linewidth]{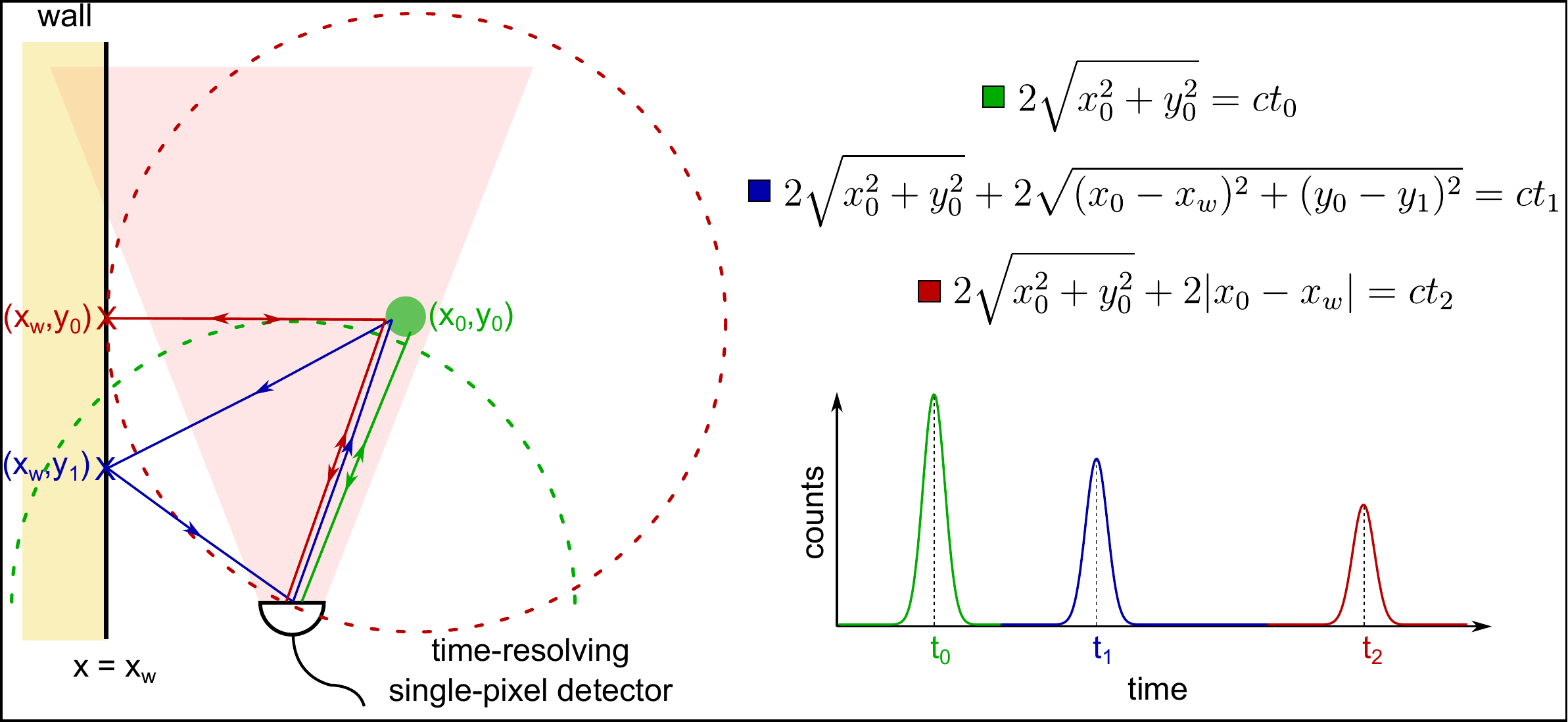}
    \caption{ Locating an object with time data and multipath echoes. By having some information about the object's environment, for instance the arrangement of a wall with respect to the emitter/detector, one can identify the object's coordinates with a simple set of equations and measurements of the arrival time of the pulsed waves that follow paths with 1, 2, and 3 bounces (presented in the form of a time histogram here). See text for further details.}
    \label{fig:echolocation}
\end{figure}

We discuss first the following question: is it possible to estimate the position of an object by just looking at the wave echoes recorded with a time-resolving single-point sensor? Using two or more sensors would make the problem trivial by simply using triangulation and direct reflections (first echo) from the object \cite{faccio:2017:oe}. However, restricting the problem to a single detector paired with unstructured illumination makes the problem ill-posed. We will show that this problem can be solved provided that we have access not just to the first echo from the object but also to secondary echoes originating from multiple bounces between the object and the room walls. Let us consider the simplest possible scenario in 2D, as depicted in Fig.~\ref{fig:echolocation}.

The emitter and the detector are both located at the origin of coordinates, for simplicity. The scene is flash-illuminated with pulsed waves emitted within an equilateral triangle (translucent red), while we assume a point detector receives any wave incident at any angle and retrieves time-of-flight (ToF) information with a temporal resolution given by the a temporal impulse response function (IRF) that we can also control. We consider a scene only consisting of a mirror-like wall (perfect reflection) and a point-like scatterer at position ($x_0,y_0$) (i.e. we restrict to the 2D case), acting as our object of interest. Also for simplicity, we assume that the distance from the wall to the detector $x_w$ is known. The scatterer (depicted with a green circle) is isotropic, i.e. it reflects waves in all directions within a circle. Our aim is to determine the position of the scatterer ($x_0,y_0$) from the wave echo (temporal information) without ambiguity by using as many bounces as needed. 

By bounces we mean reflection events which a pulse wave suffers upon propagation. For instance, first wave echoes bounce only once and return to the detector. This is the case of direct reflections from the scatterer, shown with green arrows in Fig.~\ref{fig:echolocation}. Waves can also bounce first on the scatterer, then on the wall, and then return to the sensor; this would be a 2-path case, shown with blue arrows in Fig.~\ref{fig:echolocation}. 3-path wave echoes correspond to waves that reflect first on the object, then on the wall, return to the object, and from there to the detector (shown with red arrows in Fig.~\ref{fig:echolocation}). As we assume specular reflection from the wall, these three cases described above are geometrically the only possible ones (apart from repetitions of object-wall reflections) where the waves will be reflected back to the detector, resulting in a time histogram similar to the one in Fig.~\ref{fig:echolocation}. Note that the peaks on the time histogram have decreased amplitude for an increasing number of reflections because of non-perfect reflectivity ($R<1$), as well as some finite width determined by the IRF of the detection system. 1-path events are necessarily the ones with smaller time-of-flight and therefore they will correspond to the peak at $t_0$. With simple geometry, we then obtain that: 
\begin{equation}
2\sqrt{x_0^2+y_0^2} = c t_0,
\label{eq1}
\end{equation}
which indicates that there are an infinite number of positions placed on a circle (depicted with green dashes in Fig.~\ref{fig:echolocation}) where the object can be. 
2-path events, corresponding to $t_1$ on the time histogram, provide further insight on the objects' position:
\begin{equation}
\sqrt{x_0^2+y_0^2} + \sqrt{(x_w - x_0)^2+(y_1 - y_0)^2} + \sqrt{x_w^2 + y_1^2} = c t_1,
\label{eq2}
\end{equation}
at the cost of introducing a new unknown, $y_1$, that does not allow us to place the object on the circle described by Eq.~(\ref{eq1}). 
Looking at the information that the 3-path event (corresponding to $t_2$) provides, we obtain the following identity:
\begin{equation}
2\sqrt{x_0^2+y_0^2} + 2|x_0-x_w| = c t_2,
\label{eq3}
\end{equation}
which can be used together with Eq.~(\ref{eq1}) to locate the object. Eq.~(\ref{eq2}), corresponds to a new circle (red dashes) intersecting with the point on the wall at $(x_w,y_0)$ and compatible with the distance $|x_0-x_w|$ and the measurement $t_2$. The origin of this circle coincides with the coordinates of the object. Therefore, by placing the origin of this new circle (or solving the system of equations given by Eqs.~(\ref{eq1}) and (\ref{eq3})) on the green circle described by Eq.~(\ref{eq1}) allows obtaining $(x_0,y_0)$, which demonstrates that the scatterer position can be unequivocally determined with a single detector by simply using multipath time echoes up to 3 bounces, under some assumptions.

\subsection{Physical model of our numerical simulations}
\label{physical_model}
\begin{figure}[h]
    \centering
    \includegraphics[width=1\linewidth]{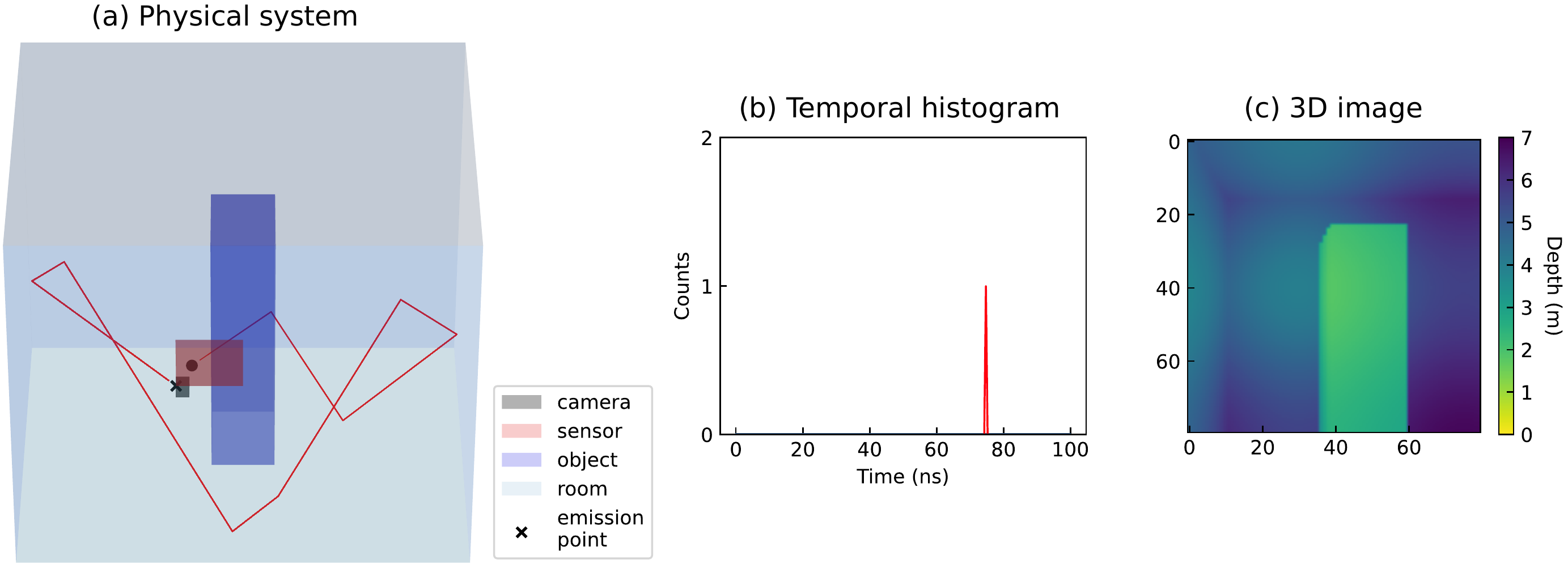}
    \caption{Physical system for our model for the 3D numerical simulations. (a) 3D visualisation of our physical system: a rectangular cuboid (blue) moves within a room. Pulsed waves are emitted at a random angle within azimuth and elevation in $[67.5^{\circ},67.5^{\circ}]$. The red line shows one example of the multipath reflections of a pulse, which eventually reaches the detector (red plane) that records its arrival time. (b) Collecting a high number of pulses allows creating a time histogram. (c) Colour-depth encoded 3D view of the scene.}
    \label{fig:model}
\end{figure}

We use numerical simulations based on a physics-inspired model to test the validity of this multipath echo-based imaging approach, see Fig.~\ref{fig:model} and Supplementary Video 1 \cite{SV}. 

Our scene consists of a closed room, which is a hollow rectangular cuboid, i.e. with $X, Y, Z$ dimensions that are not necessarily equal. The room has uniform walls, ergo they interact with the waves all in the same way. Inside this room, we place an object (a smaller rectangular cuboid) that can move freely via translation along the floor-plane. We also consider an emitter of pulsed waves and a planar time-resolving single-pixel detector. We consider that the probe pulse is emitted in all directions within azimuth and elevation angles $\theta$ and $\phi$, respectively. In other words, we assume that the probe pulse can be decomposed spatially as the sum of a high number of individual pulsed plane-waves with a $k$-vector angle within $[-\theta,\theta]$ and $[-\phi,\phi]$. Each of these pulsed waves travel within the scene at a fixed speed, $c$, and are reflected by the room walls and the object. 

Eventually, the waves hit the single-pixel detector, which acts like a stop-watch and provides the arrival time of the pulses. The probe pulses are emitted at a specific rate, $\eta$, which determines when the single-pixel detector timer is reset to zero: only echo waves returned within a time $1/\eta$ contribute to the time histogram. Non-perfect reflectivity losses in walls and objects can be introduced via a reflectivity factor $R$ that provides the probability of the wave to be reflected, such that after $N$ bounces, the probability that the wave has not been absorbed is $(1-R)^N$. Although reflectivity can also depend on the angle of incidence of the wave with respect to the surface, we have not included this in our simulations.

Therefore, in practice the reflectivity of the materials at each wave regime limits the maximum number of bounces that the waves suffer before being absorbed. We also consider the diffusion of waves when they hit the scene walls and objects via a specularity factor, $s$, that accounts for the scattering solid angle at which waves can be reflected. Mirror-like surfaces have a scattering angle of 0~rad ($s = 1$), while so-called Lambertian surfaces have a scattering angle of $\pi\,\rm{rad}$ ($s = 0$). Surfaces between these two extreme cases are considered as glossy. Optical waves have $s \ll 1$, while GHz electromagnetic (e.g. RADAR) and kHz acoustic waves (both mm-wave) have typically $s \lesssim 1$, especially for the cases considered here, where the probe pulse wavelengths are much smaller than the dimensions of the objects \cite{fuschini:2017,goulianos:2017,neubauer:1986}. Therefore, our physical model only distinguishes between waves at different regimes (e.g. optical, radio, or acoustic) via the reflectivity and specularity factors. 
%

\subsection{Simulations}
\label{simulations}
Here we describe the numerical simulations and results we obtained based on the physical model, for a room with dimensions $4\,\rm{m}\times7\,\rm{m}\times7\,\rm{m}$, with smooth, mirror-like walls of reflectivity $R=1$ and specularity $s=1$.
In the room, we place a point-source emitter of probe pulses placed at $(0.5,-1,0.5)$ emitting waves within azimuthal and elevation angles $\phi$ and $\theta$, both within $[-67.5^{\circ},67.5^{\circ}]$, at a pulse repetition rate $\eta = 10\,\rm{MHz}$. Next to the emitter, we placed a flat detector with dimensions $1\,\rm{m}\times1\,\rm{m}$ in the $YZ$ plane. In this room, we place a rectangular cuboid object with dimensions $1\,\rm{m}\times1\,\rm{m}\times5\,\rm{m}$, with reflectivity $R=1$ and specularity $s = 1$, which moves in the room by being translated in the $XY$plane such that its bottom is at $Z = 0$.

We control the number of allowed reflections per emitted pulse, which allows us to study the quality of the retrieved images for some set number of reflections. Every time histogram is obtained by emitting 10,000 pulsed plane waves per object position, following the physical model described above, and measuring the time-of-flight of the waves that return to the sensor. In order to speed up our simulations, we cropped out any pulses whose round trip time of flight would have been longer than the time between 2 pulses, or $100\,\rm{ns}$ - in practice this meant ignoring some portion of waves that reflected 8 or more times, as these histograms natively contained photons with $>100\,\rm{ns}$ arrival times.

To generate a simulated 3D image, we scanned the room from the camera position $[0.5,-1,0.5]$ over azimuth $\theta$ within $[-60^{\circ},100^{\circ}]$ and elevation $\phi$ within $[-80^{\circ},80^{\circ}]$. The asymmetry in the azimuthal angle was designed to rotate the field of view towards the bulk of the room, away from the leftside wall, for a better perspective.
\subsection{Image retrieval algorithm}
\label{algorithm}
In order to solve the inverse problem of providing an estimate of the scene from the time histogram generated by the waves echoes, we use a deep neural network algorithm.

First, we assemble a database of temporal histogram - depth image pairs, which form the inputs and labels of our supervised training scheme, using the numerical simulation described above. We created 2000 data pairs for training, and 100 for testing, per maximum number of scattering events. As we simulated single- to 10-path event scenarios, this gave a total of 20,000 and 1000 input-label pairs respectively. We set aside our testing data for later, to evaluate the neural networks.

\begin{figure}[h]
    \centering
    \includegraphics[width=1\linewidth]{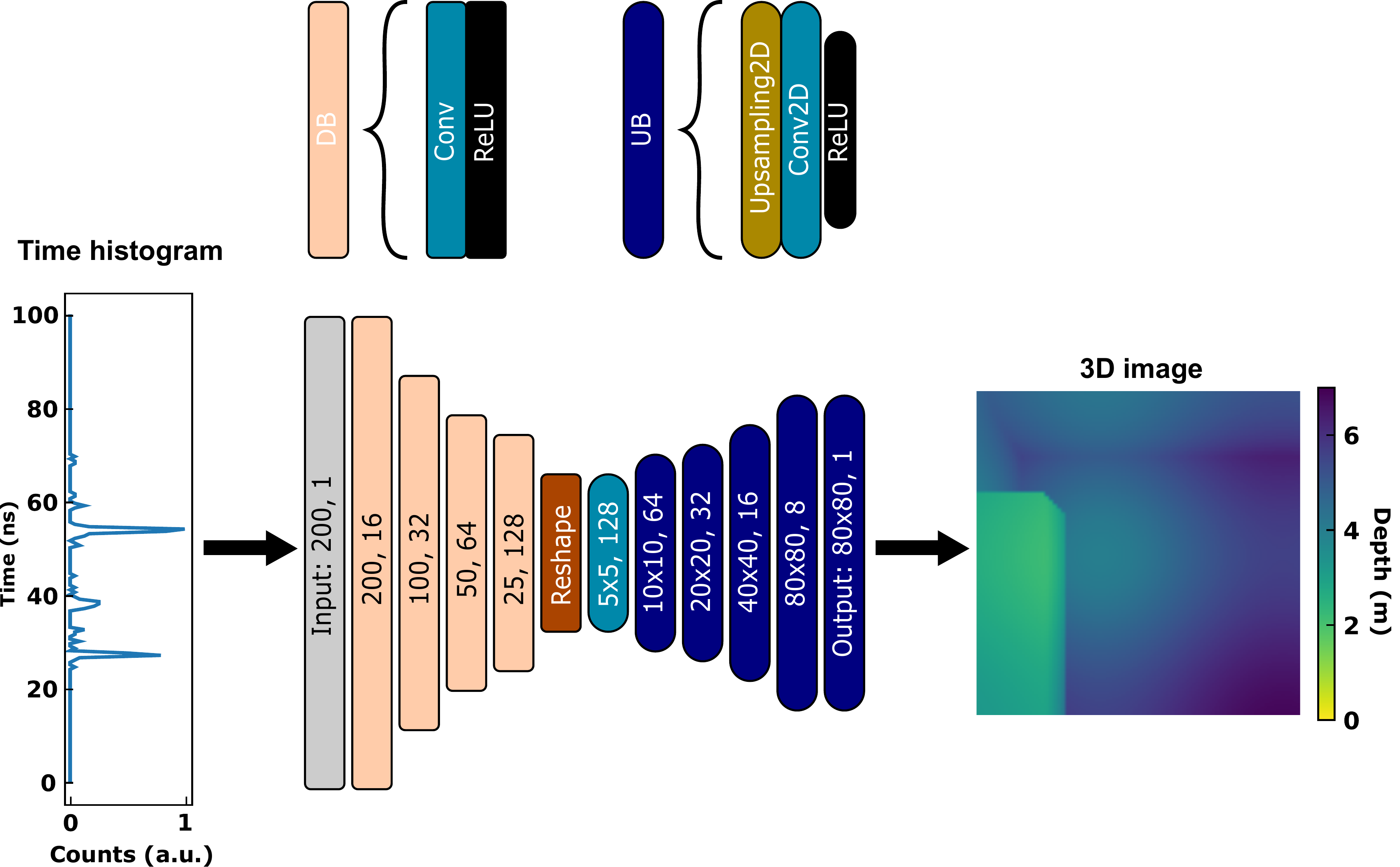}
    \caption{Deep neural network layout and operation. The input time histogram is passed through a series of downsampling blocks (DB), each of them consisting of a convolutional layer followed by a Rectified Linear Unit (ReLU) activation function. After 4 DB, we use a series of up-sampling blocks (UB), which consist of 2-D up-sampling layers, 2-D convolutional layers, and a ReLU activation function, being a depth-in-color encoded 3D image the output of the network.}
    \label{fig:NN}
\end{figure}

In the second phase, the training data pairs are used to train the network. For each of the maximum-scattering-event-number scenarios, we trained separate networks on the corresponding data pairs, albeit with the same architecture. The neural network architecture,  sketched in Fig.~\ref{fig:NN}, is fully convolutional. It has an hourglass shape such as to force information through a bottleneck, which promotes the network to compress the input into some compact representation. For our downsampling blocks (DB), we have a series of convolutional layers with kernel size 7, strides = 2 and each convolutional layer is followed by a Rectified Linear Unit (ReLU) activation function. Between the bottleneck and the output layer, there are a series of up-sampling blocks (UB), which consist of 2-D up-sampling layers, 2-D convolutional layers with kernel size $5\times5$ and strides = (1,1), and a ReLU activation function. 

We trained our neural networks using conventional machine learning methods, namely loss back-propagation via mini-batch gradient descent, implemented on a batch size of 100 using adaptive moment estimation (Adam \cite{kingma2014adam}). Our loss function was pixelwise mean-squared-error (MSE) - see Eq.\ref{Eq:MSE}. In order to prevent overtraining, we first validated the number of epochs for which to train the neural networks on 200 histogram-image pairs. In this way, we ascertained that the ideal number of training epochs increased as the maximum number of scattering events, from 110 epochs for single scattering event data to 350 for up-to-10 scattering event data. Simultaneously, the total training time of a single network increased, from ~40 to ~130 seconds respectively.

As stated in the main paper, for the RF and acoustic experiments we trained our neural networks on 9000 and 5000 temporal histogram-3D image pairs, and tested them on 865 and 500 respectively. The corresponding Supplementary Videos for our experimental results can be found in Ref. \cite{SV}. For these datasets, we implemented slightly different architectures compared to our simulated neural networks: we maintained the same form of up- and downsampling blocks, but each block had a different in- and output shape, and a different number of features. For the RF neural network, the input size (i.e. the number of bins in the temporal histogram) was 256, and the output size was $60\times80$ while the acoustic neural network was fed with inputs of size 9600, and 3D images of $64\times64$ pixels. For both the RF and acoustic neural networks, the number of features started at 64, increasing to 256 by the bottleneck, and then kept at 256 until the final layer. The increase in the number of convolutional features was chosen because the experimental scenes had a lot more variation than the simulated scenes (variety of objects, of various shapes and a range of reflectivities and specularities, uneven walls, etc.), and the corresponding histograms and images consequently showed more variability. Therefore the neural networks were designed to be able to correlate a greater number of input and output features. Otherwise, our training procedure was kept the same as for the simulated data, i.e. we used the same batch-size, loss, gradient descent optimizer, hardware, software, etc.

In the final phase of our approach and only after the deep learning algorithms are trained, we fed the latter with a single time histogram with the wave echo recording from the testing dataset, which provided a 3D image estimate of the scene.

\begin{figure*}[h]
    \centering
    \includegraphics[width=0.9\linewidth]{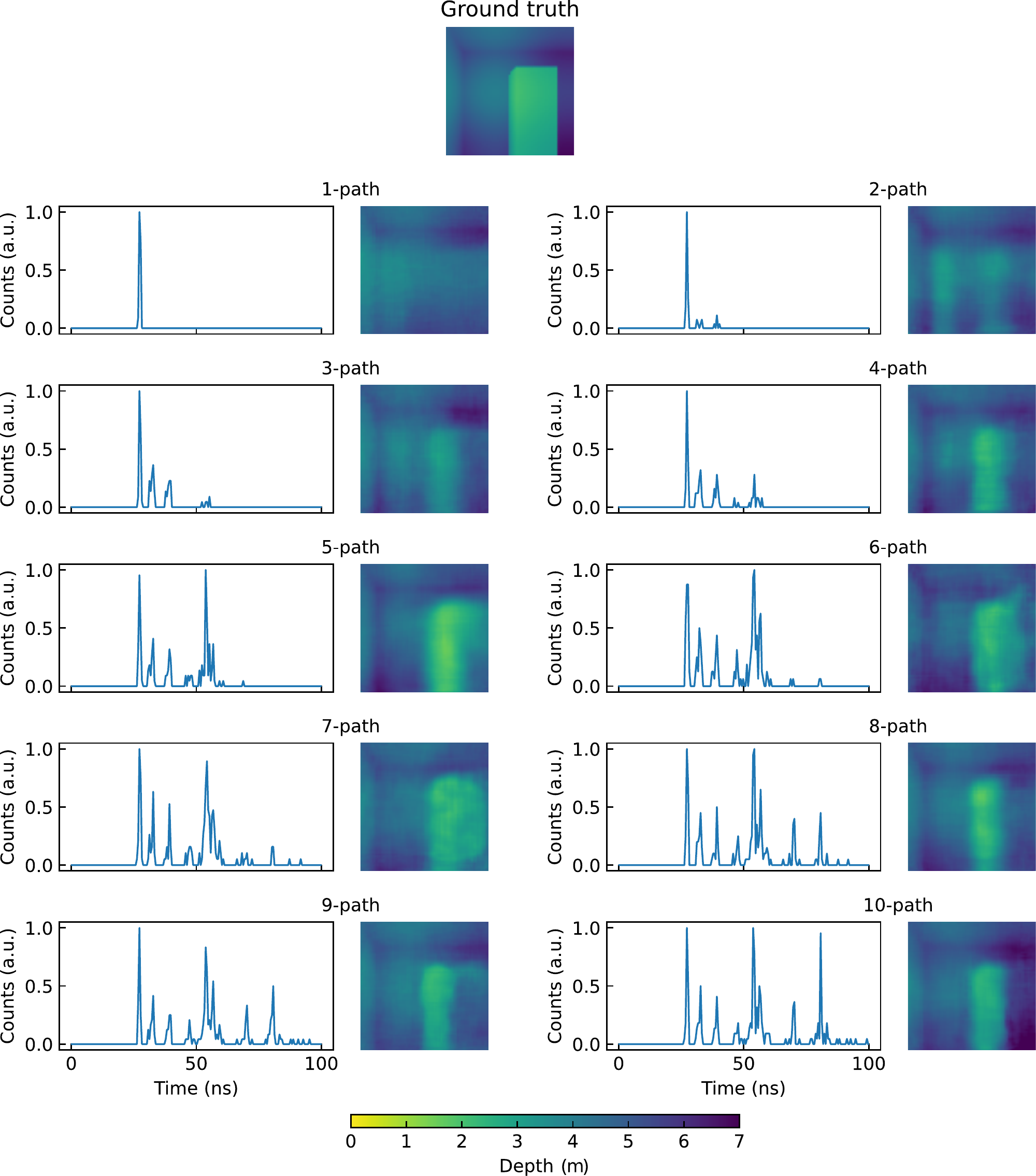}
    \caption{Image reconstruction with different multipath events. We show the time histograms obtained for 1-... 10-path events and corresponding depth-images obtained with the image retrieval algorithm. The ground truth depth-image is shown on the top.}
    \label{fig:simulations}
\end{figure*}

In Fig.~\ref{fig:simulations} we visualise the influence of the number of path events in our temporal echoes on the image retrieval, for one particular case from our test dataset (ground truth presented on the top image). As can be seen from the first and third columns, allowing waves to bounce more than once populates the time histogram in the horizontal axis, which increases its information content. This has a direct impact on the image reconstruction: the more bounces are considered, the closer the retrieved and ground truth images are. See Supplementary Video 2 \cite{SV} for reconstructions from our full test dataset. In general, it can be appreciated that 1-path events lead to ghosting in the reconstructed image, because of the degeneracy problem outlined in Eq.~(\ref{eq1}). Adding more path events clearly allows a better image reconstruction of the scene, especially when the object is placed at the sides of the image.

Training was performed in Python 3.7.9 using TensorFlow 2.1.0 \cite{tensorflow2015} and Keras \cite{keras}, on an NVIDIA GeForce RTX 2080 Ti GPU.

\subsection{Performance on unseen individuals}
\begin{figure*}[h]
    \centering
    \includegraphics[width=1.0\linewidth]{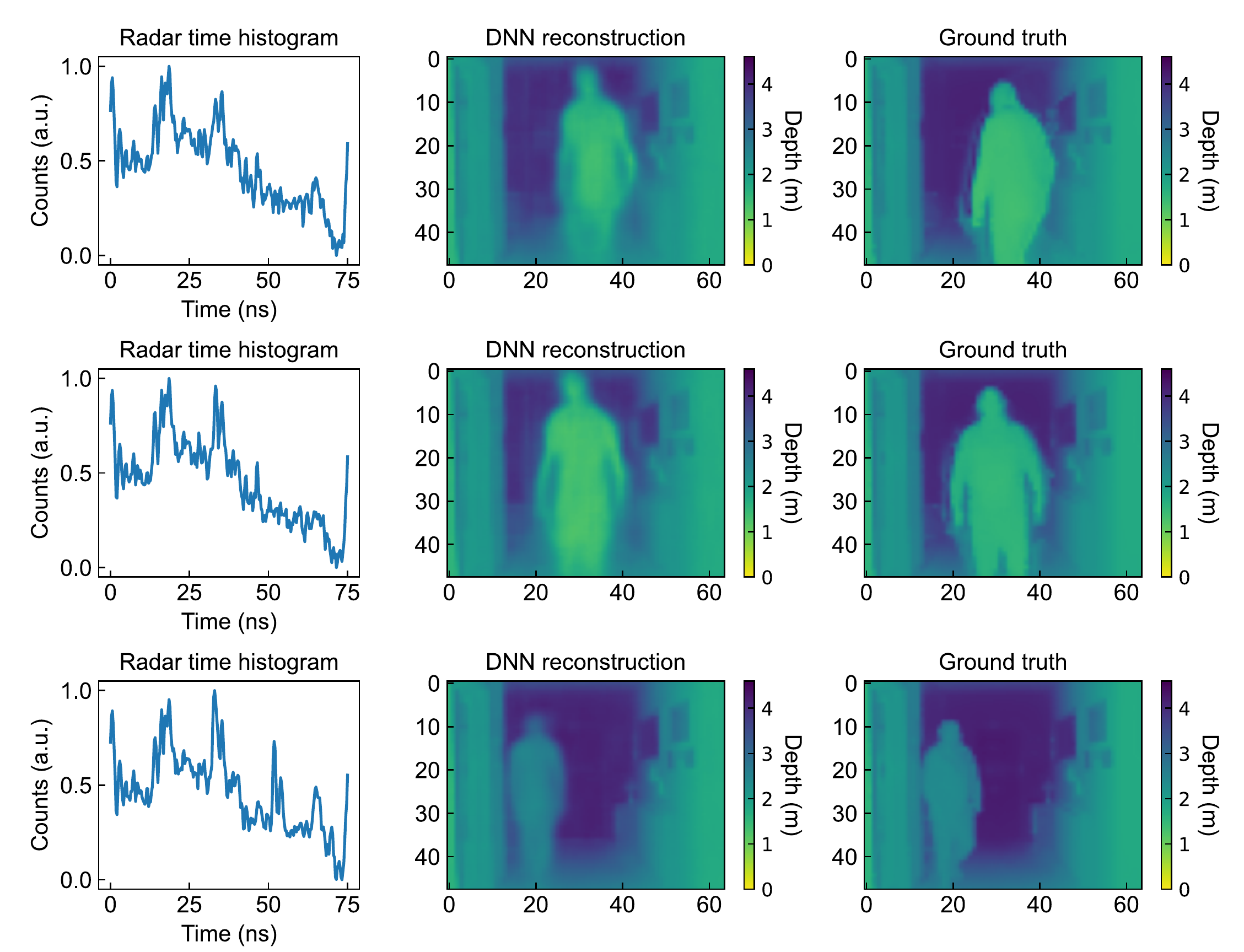}
    \caption{Performance of the technique with unseen individuals. First column shows the recorded time trace, second column shows the reconstructed image provided by the deep neural network (DNN), while the last column is the corresponding ground truth recorded with a ToF camera.}
    \label{fig:unseenind}
\end{figure*}

It is an interesting question to address whether our technique can be trained just with one individual and tested on different individuals. To answer this question, we conducted an experimental test where we trained the image retrieval algorithm with data gathered using one individual, and then operated the algorithm with data from a different individual. Our results show that the algorithm is able to provide the general shape and depth of the individual (see Fig.~\ref{fig:unseenind}). This allows us to successfully operate our technique on different individuals, while training only on one.

\subsection{Performance metrics}
\label{performance_metrics}
Our reconstruction-quality metrics of choice are the pixelwise mean-squared-error (MSE) between reconstruction and ground truth, and the intersection over union (IOU) of the foregrounds of the reconstruction and ground truth.
Formally, for reconstruction image $R$ and ground truth image $G$, both of dimensions $M\times N$, containing pixel values $R_{ij}$ and $G_{ij}$ respectively, we can write MSE as:

\begin{equation}\label{Eq:MSE}
    MSE = \frac{1}{N\times  M}\sum_{i=1}^{M}\sum_{j=1}^{N}(R_{ij}-G_{ij})^2
\end{equation}

Having a low mean squared error means that on average, the depth value predicted at a random pixel value matches the true depth value well.

\begin{figure*}[h]
    \centering
    \includegraphics[width=0.65\linewidth]{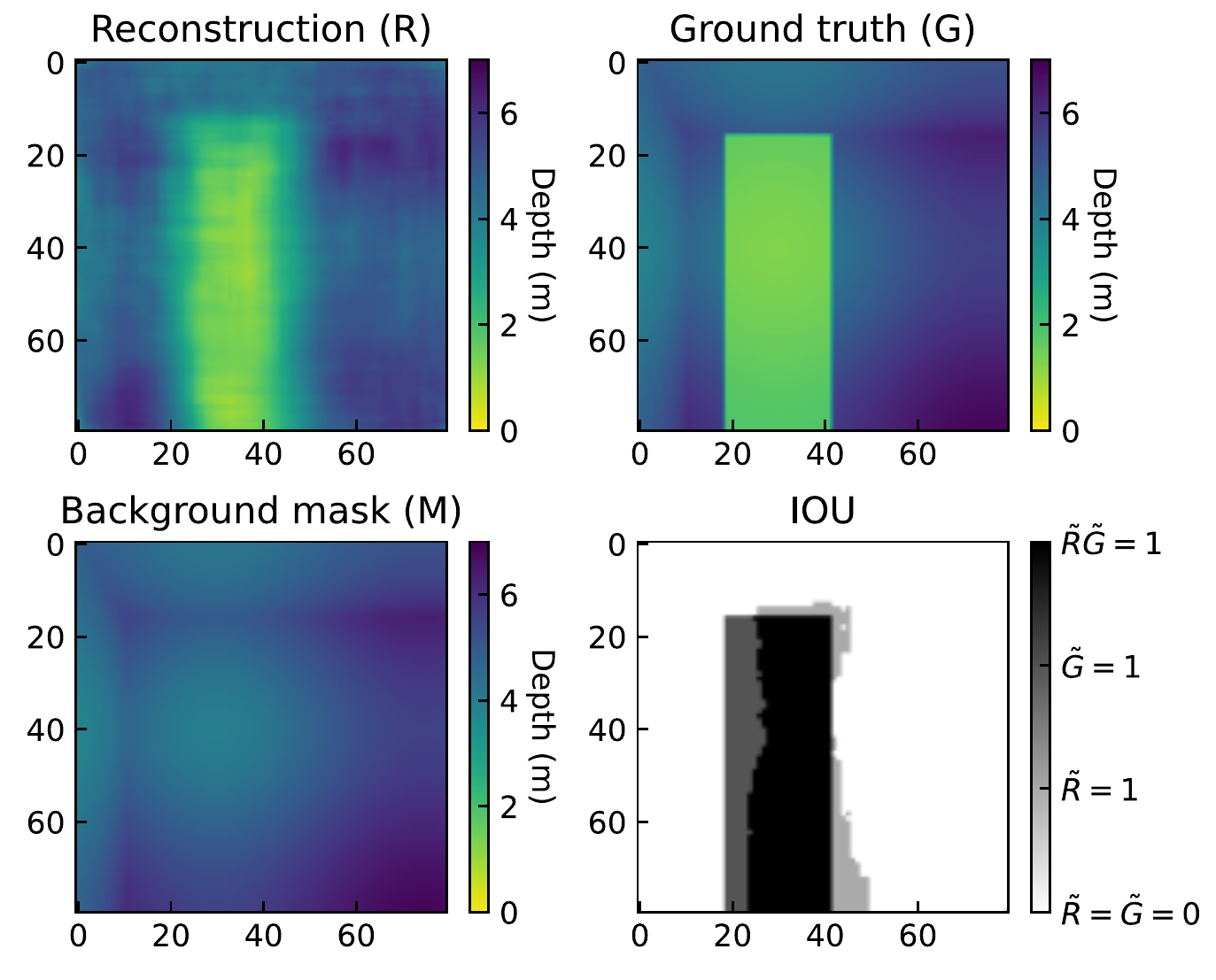}
    \caption{Visualisation of the meaning of the metric intersection over union. In the top left, we have a poor reconstruction, in the top right, its corresponding ground truth. We compare $R$ and $R$ to the mask shown in the bottom left, to identify the foreground regions. These regions, denoted as $\tilde{R}=1$ and $\tilde{G}=1$, are shown in light and dark grey respectively. Then, their intersection is the area of the region shown in black on the bottom right, and the union is the combined area of everything that is not white. Finally, the IOU is found according to Eq.~(\ref{Eq:IOU}).} 
    \label{fig:iou}
\end{figure*}

Intersection over union is a less common metric, which focuses on how well the shapes of the foreground objects are reconstructed, and is invariant to changes in the relative size of the foreground object within the field of view. To calculate IOU, we take a background mask, and compare our ground truth and prediction images to this mask; with this comparison, we binarise the ground truth and prediction images into foreground (variable) and background (static) pixels. This is shown in Fig.~\ref{fig:iou}, in the bottom-right in greyscale. Their intersection, then, is simply the area of overlap, while the union is the combined area of all non-background regions. 

We can write this mathematically as follows: for background mask pixels $M_{ij}$, we binarise our reconstruction pixels $R_{ij}$ and ground truth pixels $G_{ij}$ to obtain $\tilde{R_{ij}}$ and $\tilde{G_{ij}}$, where $i$ and $j$ are in used the same notation as in Eq.~(\ref{Eq:MSE}), as such:
\begin{equation}\label{Eq:binarise}
\begin{aligned}
\tilde{R_{ij}} = \begin{cases} 
       1 & (M_{ij}-R_{ij}) \geq \kappa r\\
       0 &(M_{ij}-R_{ij}) < \kappa r
   \end{cases}\\
\tilde{G_{ij}} = \begin{cases} 
       1 & (M_{ij}-G_{ij}) \geq \kappa g\\
       0 &(M_{ij}-G_{ij}) < \kappa g,
   \end{cases}
\end{aligned}
\end{equation}
where $r$ and $g$ are the maximum pixel-wise differences between the mask and the reconstruction, and the mask and ground truth, respectively. The threshold $\kappa$ was set at $0.5$, $0.2$ and $0.2$ for the synthetic, RF and acoustic data respectively, as these values were found to consistently give visually good binarisation.
Then, IOU is:
\begin{equation}\label{Eq:IOU}
    IOU =\frac{\sum_{i=1}^{80}\sum_{j=1}^{80}(\tilde{R_{ij}}\tilde{G_{ij}})}{\sum_{i=1}^{80}\sum_{j=1}^{80}\max\{\tilde{R_{ij}},\tilde{G_{ij}\}}}.
\end{equation}

A large intersection over union means the neural network reconstructs the foreground object largely at the same position as where it was in the ground truth image.

\begin{figure}[h]
    \centering
    \includegraphics[width=1\linewidth]{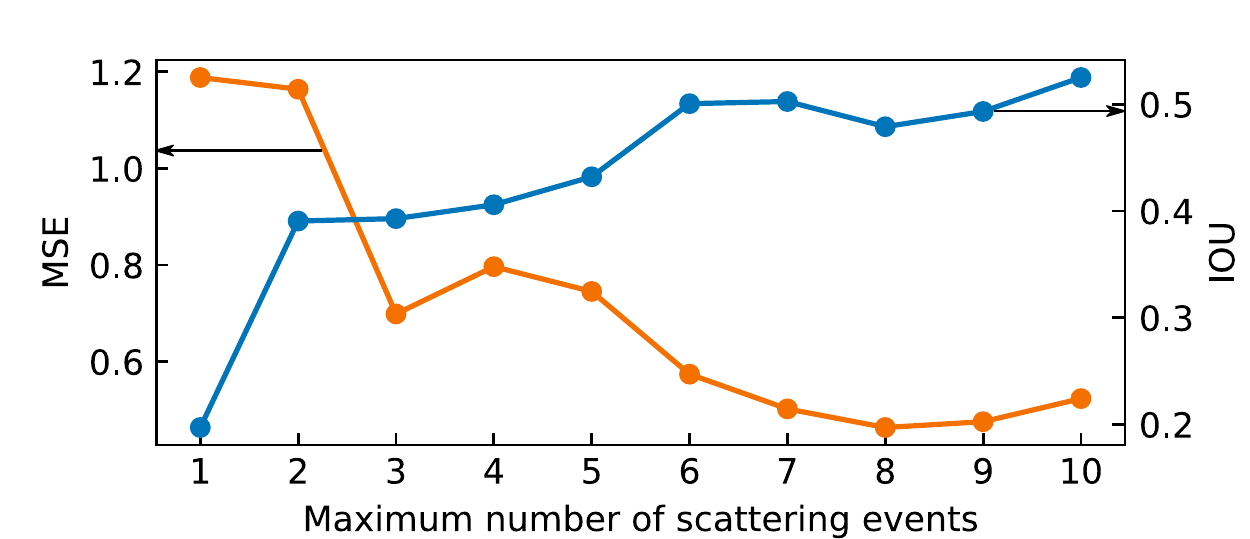}
    \caption{Mean mean-squared-error (MSE) and intersection over union (IOU) obtained from our 10 model replicas, each averaged over 100 histogram-3D image pairs, with increasing number of maximum scattering events.}
    \label{fig:metrics_comp}
\end{figure}

In Fig.~\ref{fig:metrics_comp} we compare the evolution of MSE and IOU as a function of the maximum number of scattering events, for our 100 testing 3D image-neural net reconstruction pairs. As stated in the main text, we trained 10 neural networks for each of the maximum scattering event scenarios, and averaged over the MSE and IOU obtained from the predictions of these 10 ANN copies on the test set. This was done to minimise the specificity of our predictions, and correspondingly, our performance metrics, on the starting configurations of our ANNs. 

Clearly, adding more bounces enhances the ability of the algorithm to retrieve images, which demonstrates the advantage of using waves that are not absorbed after the first bounce for such an imaging scheme. MSE, the loss metric of our neural network, improves up until 8 scattering events, and IOU up until 10. Neither MSE nor IOU improve homogeneously, however the local minima/maxima are probably attributed to the inherent randomness of our simulator as opposed to actual local maxima in the amount of information in the histograms.

\subsection{Information theory analysis}
\label{informaton}
To quantify the gain in information when including an increasing number of bounces, we use the principle of Shannon entropy, joint entropy and mutual information derived from information theory \cite{shannon, MacKay_info, elements_of_info}. Information is formally defined as the number of bits required to describe a reduction in uncertainty when observing a random variable $X$ on a set $\mathcal{X}$ at a value $x\in \mathcal{X}$ that occurs with probability $p(x)$. This can be calculated by $-\log_2{p(x)}$ and is commonly referred to as self-information or Shannon information. Furthermore, the expectation value of uncertainty reduction when observing variable $X$ is known as the Shannon entropy:
\begin{equation}\label{Eq:entropy}
    H(X) = -\sum_{i=1}^N{p(x_i)\log_2{p(x_i)}}.
\end{equation}
We first quantify the information of temporal data containing a signal due to a single bounce. In this context, we define the time of arrival of a photon as a random variable and neglect the variable number of counts at a given time-bin to simplify the calculation. Given 2000 examples of single bounce histograms from the stochastic model described above, we find identical temporal traces within the data; let us call the set of all unique traces $\tilde{\mathcal{X}}$. We then re-assign these traces to a new variable $\tilde{X}$, which can assume an integer label $\tilde{x}\in \tilde{\mathcal{X}}$. After normalising the distribution of $\tilde{X}$ to a probability mass function, we then calculate entropy $H(\tilde{X})$ by Eq.~(\ref{Eq:entropy}).
To assess what additional information is gained by including photons from a second bounce in the temporal-trace, we use the joint entropy $H(\tilde{X},\tilde{Y})$ of the distribution of identical temporal-data for a single bounce $\tilde{X}$, and the equivalent distribution for data containing up to two bounces $\tilde{Y}$. The joint entropy is simply the Shannon entropy for a joint distribution of $\tilde{X}$ and $\tilde{Y}$ on $\tilde{\mathcal{X}}$ and $\tilde{\mathcal{Y}}$ respectively, and describes the expected uncertainty reduction when observing $(\tilde{x}, \tilde{y})$ where $\tilde{x}\in \tilde{\mathcal{X}}$ and $\tilde{y}\in \tilde{\mathcal{Y}}$. For context, the uncertainty reduction of observing a given temporal trace containing photons from one bounce can be increased when also considering the data from the second bounce. We find the joint probability distribution by the relation $p(\tilde{x},\tilde{y}) = p(\tilde{y}|\tilde{x})p(\tilde{x})$ and use this to calculate joint entropy by:
\begin{equation}\label{Eq:joint_entropy}
    H(\tilde{X},\tilde{Y}) = -\sum_{j=1}^M{\sum_{i=1}^N{p(\tilde{x}_i, \tilde{y}_j)\log_2{p(\tilde{x}_i,\tilde{y}_j)}}}.
\end{equation}
We then repeat the Joint Entropy calculation to compare the data containing $<n$ bounces and $<(n+1)$ bounces, and present the gain in uncorrelated information when including photons from additional bounces using the definition of mutual information $MI(\tilde{X};\tilde{Y})$ which describes the information shared by two random variables due to correlation:
\begin{equation}\label{Eq:mutual_info}
    MI(\tilde{X};\tilde{Y}) = H(\tilde{X})+H(\tilde{Y})-H(\tilde{X},\tilde{Y})
\end{equation}
We rearrange Eq.~(\ref{Eq:mutual_info}) to find the additional uncorrelated information in $\tilde{Y}$, i.e. the mutual information $MI(\tilde{X};\tilde{Y})$ subtracted from the total information in $\tilde{Y}$ or equivalently, $H(\tilde{X},\tilde{Y})- H(\tilde{X})$.
This gain in \textit{uncorrelated} information $UI(\tilde{X};\tilde{Y})$ increases as photons which have experienced an increasing number of bounces are included in the data as shown in Fig.~2(a) of the main document. 
Clearly, there is a dramatic increase in information when including photons which have experienced two bounces compared with only a single bounce, however photons experiencing five bounces or more show no increase in information content for this data set. It is expected that any information about the 3D scene given by the knowledge of the number of counts at a given time-bin may further contribute the reconstructed image quality. 

The neural network algorithm used to reconstruct the 3D images can leverage both the time of arrival and the number of counts which may account for improvement of reconstructions when given data with photons experiencing five or more photons.
If we consider the 3D image to be an input from an information theory point of view and the histogram to be the output, we hypothesise that adding more bounces beyond the actual Shannon entropy increases the redundancy of the transmitted data i.e. is similar to a better encoding scheme. This reduces the probability of a bit error when propagating down a noisy channel, in this case the stochastic sampling and discretisation performed at the radar/microphone. 
This hypothesis can be illustrated with the mirror room image from Fig.~2(b) in the main document: adding more bounces creates replicas of existing mirror images. These replicas do not add more information in a lossless/noiseless environment, but when the environment is sampled in a noisy way, the added redundancy can help in localising and identifying the object.

\end{document}